# Cultivating Long-Term Planning, Collaboration, and Mission Continuity in Astrobiology Through Support of Early Career Researchers

A white paper for the 2025 NASA Decadal Astrobiology Research and Exploration Strategy (DARES)

**Authors**: Elizabeth Spiers[1,6] (*elizabeth.spiers@whoi.edu*), Jessica Weber[2] (*jessica.weber@jpl.nasa.gov*), Katherine Dzurilla[2] (*katherine.a.dzurilla@jpl.nasa.gov*), Erin Leonard[2] (*erin.j.leonard@jpl.nasa.gov*), Sierra Ferguson[3] (*sierra.ferguson@swri.org*), Natalie Wolfenbarger[4] (*wolfenbarger@lanl.gov*), Kristian Chan[5] (*kristian.chan@jhuapl.edu*), Perianne Johnson[6] (*perianne.johnson@jsg.utexas.edu*), Kirtland Robinson[7] (*kirtland.robinson@asu.edu*), Chase Chivers[8] (*chase.j.chivers@dartmouth.edu*)

[1]Woods Hole Oceanographic Institution; [2]NASA Jet Propulsion Laboratory, California Institute of Technology; [3]Southwest Research Institute; [4]Los Alamos National Laboratory; [5]John Hopkins APL; [6]University of Texas at Austin; [7]Arizona State University; [8]Dartmouth College

---

**Community Endorsements:** *Taylor Plattner, tplattner24@gatech.edu; Kevin T. Trinh, kttrinh1@asu.edu; Jasmine Singh, jasmine.singh@my.utsa.edu; Emily Paris, erparis@stanford.edu; Pilar Vergeli, pvergeli@asu.edu; Steffen Buessecker, sbuessecker@stanford.edu; Nuri Park, npark13@asu.edu; Ashley M. Hanna, Ahanna13@umd.edu; Morgan Cable, Morgan.l.cable@jpl.nasa.gov; Ilankuzhali Elavarasan, ilankuzhali.elavarasan01@utrgv.edu; Alta Howells, alta.howells@colorado.edu; Sabrina Elkassas, elks@mit.edu; Caroline Haslebacher, caroline.haslebacher@contractor.swri.org; Jordan McKaig, jmcakig3@gatech.edu; Lea Adepoju, ladepoju6@gatech.edu; Snigdha Nellutla, snellutla3@gatech.edu; Amanda Stockton, astockto@gatech.edu; Chad Pozaryck, chad.ian.poz@gmail.com; Biya Haile, bhaile3@gatech.edu; Christopher E. Carr, cecarr@gatech.edu; Anthony Burnetti, anthony.burnetti@biosci.gatech.edu; Vahab Rajaei, vrajaei3@Gatech.edu; Catherine Fontana, catherine.fontana@colorado.edu; Christina Buffo, cbuffo6@gatech.edu; Ellie Hara, harae@rpi.edu; Gabriella Rizzo, grizzo3@huskers.unl.edu; Jorge Coppin-Massanet, jyc77@cornell.edu; Veronica Hegelein, vah38@cornell.edu; Charlie Detelich, ced237@cornell.edu; Sara Miller, sgm96@cornell.edu; Mariam Naseem, mnaseem@umd.edu; Lily Clough, Lily.a.clough@nasa.gov; Srishti Kashyap, srishti.kashyap@colorado.edu; Laura Rodriguez, lrodriguez@lpi.usra.edu; Christopher Glein, cglein@swri.edu; Jacob Buffo, jacob.j.buffo@dartmouth.edu; Alexia Kubas, ak2248@cornell.edu*

> **The 2025 Astrobiology strategy should not only prioritize Early Career Researchers (ECRs), but it should include specific recommendations for training, support, & retention of ECR talent and workforce in Astrobiology.**

The future stability and success of NASA Astrobiology is built on Early Career Researchers (ECRs), defined as students and individuals <10 years post-PhD. They hold a central role in future planning and activities within NASA Astrobiology. Yet, the 2015 Astrobiology Strategy only mentioned graduate students twice and had no mention of postdocs or early career researchers. In this document, we identify two goals for NASA Astrobiology that can be addressed through ECR support and initiatives. We additionally list specific recommendations for programs and initiatives that would help achieve these goals, Table 1.

**Table 1: Goals and Recommendations for Early Career Researcher Training, Retention, and Support.** *Further details on this list of recommendations can be found at the end of the document*

| | |
|---|---|
| **Goal 1:** Knowledge Retention and Workforce Stability | Continued support of the NASA Astrobiology Postdoctoral Program |
| | Continued support for ECR involvement and training for missions |
| | Provide Astrobiology Graduate fellowships that are explicitly interdisciplinary |
| | Incentivize ECR funding and mentorship within data analysis programs (DAP) |
| **Goal 2:** Foster Collaboration & Strengthen Community | Continued support of research travel funds such as the Lewis and Clark Award and the NASA Early Career Collaborator Award |
| | Continued support of AbSciCon and AbGradCon with the addition of initiatives to build peer networks in-person |
| | Commit Resources to Early Career groups within the RCNs and within the AGs. |

## Goal 1: Knowledge Retention and Workforce Stability

Astrobiology initiatives often span multiple decades, requiring knowledge to be passed across generations of scientists. Ensuring a stable pipeline of early career researchers (ECRs) is essential for the long-term success of NASA Astrobiology. However, this pipeline is vulnerable to fluctuations in the funding environment, particularly during gaps between active missions. To maintain continuity, mechanisms of continued investment in ECR astrobiology researchers must exist across career stages from graduate student to research scientist. In particular, providing resources to facilitate the transition from "temporary" (e.g., graduate student, postdoc) to "permanent" positions (e.g., research scientist, assistant professor) ensures that the workforce NASA has already invested in training can be retained.

> **By funding the training and support of ECRs, NASA Astrobiology can ensure the continuity of critical scientific expertise and provide workforce stability.**

**Challenges:**

- Student Funding: Broadening graduate student engagement requires financial commitment before students are accepted into programs. Currently, programs like FINESST primarily fund students who already have secured positions, favoring those who were already committed to specific projects or would have received funding regardless. In particular, it advantages individuals at R1 institutions with higher research activity and those with a history of substantial mission involvement. Institutions without resources for student recruitment can only apply once, when the student is applying to graduate school. Considering that one of the evaluation criteria within FINESST is "exhibited depth of understanding of the research topic," this puts these students at a severe disadvantage against students who can apply multiple times and can have a couple of years to mature their project and subject knowledge in a well-funded environment.

- Career Stability: The transition from graduate school to a stable, long-term research position is often precarious, marked by gaps in funding and a series of temporary positions before securing a permanent role. These issues are inherent throughout academia and are not specific to the NASA Astrobiology community, but the potential loss of workforce is enhanced by the lack of relevant research career paths within industry and the private sector, reducing the likelihood that ECRs will return to the astrobiology research workforce later on. Due to the complexity of the problem, we have refrained from providing any specific recommendations for providing funding to post-PhD ECRs beyond further support of the NPP program. **However, we believe it is pressing to have further discussions on solutions to facilitate the jump from temporary to permanent contract positions.** Some potential mechanisms include providing funds for ECRs similar to the NIH's "K99/R00 - Pathway to Independence Award" or providing mechanisms that make it easier for ECRs to apply to ROSES funding as a postdoc (non-PI status) and migrate that funding from one institution to another.

- Retention of Talent: Retaining ECRs is essential for the long-term success of astrobiology, particularly given the field's interdisciplinary nature and reliance on multi-decade mission timelines. Losing talented ECRs can result in critical gaps in expertise, disrupt knowledge transfer, and slow scientific progress. As astrobiology depends on insights from diverse fields, maintaining a stable and engaged community of researchers is vital to sustaining collaboration, innovation, and scientific excellence over time.

## Goal 2: Foster Collaboration & Strengthen Community

The inherent interdisciplinary nature of astrobiology requires the navigation of collaborative environments and the creation of a diverse network, which can be a difficult task for ECRs. By providing cross-disciplinary training opportunities to ECRs, we can train new scientists outside of traditional scientific boundaries. This will accelerate our understanding of new topics and provide expertise that can open up new opportunities for research. In addition, opportunities for peer-to-peer networking, such as AbSciCon and AbGradCon, are vital for supporting ECRs' scientific development. Building a strong peer network is crucial for retaining ECRs in academia, particularly in interdisciplinary fields like astrobiology. Strengthening community bonds within NASA Astrobiology can foster meaningful connections across disciplines and divisions. Particularly as many of the ECRs' networking opportunities were greatly limited during the COVID-19 pandemic, the continuation of these opportunities is essential to supporting ECRs' building of peer networks and capabilities for future collaboration.

> **By investing in community-building efforts, NASA Astrobiology can improve ECR retention, enhance collaboration, and ensure the field's long-term scientific excellence.**

**Challenges:**

- <u>Difficulties working across institutions, departments, and fields</u>: Astrobiology research inherently requires collaboration across diverse disciplines, yet early-career researchers often face significant challenges navigating this interdisciplinary landscape. Academic structures frequently reinforce intellectual silos, with departments prioritizing specialized expertise over cross-field integration. This can hinder communication, limit access to critical resources, and slow progress on complex research questions that demand insights from multiple fields. For early-career scientists, building connections across institutions and specialties/disciplines can be particularly difficult, as they may lack established networks or face pressure to conform to traditional disciplinary boundaries. Overcoming these barriers requires institutional support, mentorship, and deliberate efforts to foster interdisciplinary collaboration.
- <u>Effects of COVID-19</u>: Many of the current ECRs in astrobiology received their graduate or postdoctoral training during the COVID-19 pandemic, which limited their ability to attend conferences, network, and interact with peers. It is unclear what effects this may have had or will have on the long-term retention of ECR talent. While we make no specific recommendations relevant to the cohort of ECRs whose training was affected by COVID-19, recommendations that encourage community involvement and ECR retention will benefit this cohort of individuals, mitigating some of their initial isolation from the broader astrobiology community.

## Recommendations for the Training, Support, and Retention of Early Career Researchers in NASA Astrobiology

<u>Continued support for ECR involvement and training for missions</u>. ECR's mission involvement is a critical component to the continuation and success of long time-scale missions. However, opportunities for gaining mission experience as an ECR have been historically limited. Opportunities to bring in ECRs into mission environments and training have been implemented (ex. Dragonfly Student & Early Career Investigator Program, Europa ICONS, Here to Observe "H2O"). These opportunities, in addition to programs with more structured curriculums (such as the NASA Science Mission Design Schools and the NASA Astrobiology Mission Ideation Factory), provide opportunities for ECRs to gain mission experience and help the continuation of knowledge across generations of scientists.

<u>Continued support of the NASA Astrobiology Postdoctoral Program (NPP).</u> The astrobiology NPP provides critical funding opportunities to postdocs as they exit their PhD and transition to independent research positions.

<u>Provide Astrobiology Graduate fellowships that are explicitly interdisciplinary</u> with 2+ thesis supervisors from different disciplines. Although cross-divisional and cross-directorate collaboration is valued in Astrobiology research, few graduate programs are dedicated to training those capable of leading such inherently interdisciplinary projects. This gap in workforce training can be addressed by offering fellowships to supervisor teams prior to student acceptance with some flexibility for the award disbursement to enable effective recruitment. Because the existing academic culture still places emphasis on and rewards disciplinary expertise, this funding would enable the recruitment of promising students who can apply skills developed in one field to novel and innovative problem-solving approaches in another. This enables funding and recruitment of students from nontraditional backgrounds who may have fewer opportunities for funding or would otherwise have committed to a less interdisciplinary dissertation project.

<u>Incentivize ECR funding and mentorship within data analysis programs (DAP) or similar programs</u>. DAP funding enables the training of the next generation in data analysis during periods between mission funding cycles. However, because student and postdoc funding can be prohibitively costly on limited grant budgets, DAP funding is often directed toward established scientists. By providing incentives to train and support students and postdocs on DAPs, we can increase the likelihood that student support and training will be included. Incentives for student funding and mentoring can be applied to other grants as well, but we focus on DAPs in this recommendation since they are critical for mission workforce training.

<u>Continued support of travel funds such as the Lewis and Clark Award and the NASA Early Career Collaborator Award.</u> These awards enable collaboration opportunities for ECRs across institutional boundaries, enabling research that may not have otherwise been feasible.

Continued support of AbSciCon and AbGradCon with the addition of initiatives to build peer networks in-person. Opportunities specifically for peer-to-peer networking, such as AbSciCon and AbGradCon, are vital for supporting ECR's scientific development. Particularly since many of the ECR's networking opportunities were greatly limited during the COVID-19 pandemic, the continuation of these opportunities is essential to supporting ECRs' building of peer networks. The addition of initiatives or events for peer networking at pre-existing venues, such as AbSciCon, LPSC, or RCN meetings, would further enable peer-to-peer networking for ECRs.

Commit Resources to Early Career groups within the RCNs and within the AGs. Fostering better connections between ECRs and the parent organizations within NASA Astrobiology will allow for stronger community ties. While grassroots efforts (e.g., Future Leaders of Ocean Worlds (FLOW) - NOW RCN, Early Career Council - NFoLD RCN, Europa Clipper Sunrise) have made inroads for early career groups, funding is required to really increase the effectiveness of these organizations. For example, the recent establishment of the NOW RCN's FLOW fellowship program recognizes early career scientists who volunteer their time to their community and actively pursue roles to support their development as future leaders in the field. Intentional allocation of funds to support such programs ensures that the next generation of scientists can develop the network and skills necessary to lead the interdisciplinary teams critical to research in Astrobiology.